\documentclass[%
 aip,
 amsmath,amssymb,
 reprint,%
]{revtex4-1}

\usepackage{graphicx}
\usepackage{dcolumn}
\usepackage{bm}

\usepackage{siunitx}

\usepackage[utf8]{inputenc}
\usepackage[T1]{fontenc}
\usepackage{mathptmx}
\usepackage{etoolbox}

\usepackage{CJKutf8}

\makeatletter
\def\@email#1#2{%
 \endgroup
 \patchcmd{\titleblock@produce}
  {\frontmatter@RRAPformat}
  {\frontmatter@RRAPformat{\produce@RRAP{*#1\href{mailto:#2}{#2}}}\frontmatter@RRAPformat}
  {}{}
}%
\makeatother
\begin{document}
\begin{CJK*}{UTF8}{gbsn}

\preprint{AIP/123-QED}

\title{Effective and efficient modeling of the hydrodynamics for bacterial flagella}
\author{Baopi Liu (刘葆僻)}
\email{bpliu@csrc.ac.cn}
\affiliation{Complex Systems Division, Beijing Computational Science Research Center, Beijing 100193, China}

\author{Lu Chen (陈璐)}
\affiliation{College of Physics, Changchun Normal University, Changchun, Jilin 130032, China}

\author{Ji Zhang (张骥)}
\affiliation{Zhejiang Lab, Hangzhou 311100, China}

\date{\today}
\begin{abstract}
The hydrodynamic interactions among bacterial cell bodies, flagella, and surrounding boundaries are essential for understanding bacterial motility in complex environments. In this study, we demonstrate that each slender flagellum can be modeled as a series of spheres, and that the interactions between these spheres can be accurately characterized using a resistance matrix. This approach allows us to effectively and efficiently evaluate the propulsive effects of the flagella. Notably, our investigation into bacterial motility near a colloidal sphere reveals significant discrepancies between results derived from the twin multipole moment and those obtained through resistive force theory. Consequently, neglecting the hydrodynamic interactions among cell bodies, flagella, and colloidal spheres may lead to substantial inaccuracies. Our model simplifies bacteria into a series of spheres, making it well-suited for examining bacterial motility near spherical boundaries, as well as the nonlinear deformation dynamics of elastic flagella.
\end{abstract}
\maketitle

\end{CJK*}
\section{Introduction}
\hangafter=-1\hangindent=19pt\noindent

Bacteria are ubiquitous in natural environments, and their motility mechanisms represent a critical area of investigation in both microbiology and hydrodynamics \cite{Lauga2016,Guasto2012}. Most bacteria propel themselves in fluid environments by rotating or beating their flagella, which are elongated, flexible protein appendages extending from their cell bodies \cite{Silverman1974,Lauga2009}. The characteristics of the surrounding fluid significantly affect bacterial motility \cite{Kargar2014,Petroff2015,Chang2018,Sipos2015,Kamdar2022,Spagnolie2023}. Understanding these mechanisms across various environments is essential, not only for fundamental biological research, such as fertilization \cite{Tung2021,Raveshi2021} and biofilm formation \cite{Khalid2020,Vissers2018}, but also for practical applications in bioengineering, including the development and control of artificial microswimmers \cite{Volpe2011,Takagi2014}.

Despite extensive research on bacterial motility in viscous fluids, challenges persist in the computational methods of hydrodynamics. Bacterial motility can be understood as a result of various force interactions among the flagella, cell bodies, and the surrounding fluid. In this context, viscous forces are generally greater than inertial forces, leading to a linear relationship between kinetics and kinematics \cite{Lauga2009,Happel2012,Kim2013}. Therefore, to effectively analyze bacterial motility in viscous environments, it is crucial to compute the resistance matrix for both the cell bodies and the flagella. Current approaches for calculating this matrix include resistive force theory (RFT) \cite{Gray1955}, Lighthill slender body theory (LSBT) \cite{Lighthill1976,Koens2018,Mori2020}, the singularity method \cite{Dabros1985}, the regularized Stokeslets method (RSM) \cite{Cortez2001,Cortez2005}, the boundary element method (BEM) \cite{Smith2009}, the immersed boundary method (IBM) \cite{Peskin2002,Olson2013}, and the Rotne-Prager-Yamakawa (RPY) approximation \cite{Wajnryb2013,Zuk2014,Vizsnyiczai2020}. However, each of these methodologies has inherent limitations, and different models and boundary conditions require distinct computational techniques, which can vary significantly \cite{Vizsnyiczai2020,Di2011,Temel2013,Shum2015,Pimponi2016,Shum2010}. For example, when bacteria swim near a flat surface, the hydrodynamic interactions between the flagella and the surface significantly influence the results \cite{Shum2015,Pimponi2016,Shum2010}. Recently, the investigation of bacterial motility near colloidal spheres has emerged as a prominent research area \cite{Kargar2014,Chang2018,Sipos2015,Kamdar2022}, which requires careful consideration of the hydrodynamic interactions among cell bodies, flagella, and colloidal spheres. Unfortunately, existing methods often prove inadequate for this scenario, highlighting the urgent need for developing novel computational techniques \cite{Kargar2014,Chang2018}.

Filamentous structures, such as polymers, are commonly represented using bead-spring models \cite{Loi2011,Martin2019}. In this study, we utilize Escherichia coli as a model organism, representing the flagellum as a helical arrangement of spheres \cite{Reigh2012}. This approach simplifies the computation of the resistance matrix for the flagella to that of a series of spheres. We employ the twin multipole moment (TMM) \cite{Jeffrey1973,Jeffrey1984,Durlofsky1989} to calculate the resistance matrix of the multi-sphere system. This method captures both near- and far-field interactions, providing a more accurate description of the hydrodynamic interactions between spheres. The resistance matrix derived from the TMM can be utilized to simulate the dynamics of colloids and the kinematics of active matter in shear flows \cite{Pedley1992,Guan2024}. To validate the effectiveness of our model, we conduct numerical simulations of the axial thrust force and torque generated by a single rigid helical flagellum, comparing the results with existing laboratory measurements \cite{Rodenborn2013}. Our findings indicate that the proposed model is effective for calculating the dynamics of the flagella in viscous fluids. Furthermore, this model is applicable for investigating bacterial motility on spherical surfaces \cite{Kargar2014,Chang2018,Sipos2015,Kamdar2022} and the nonlinear deformation dynamics of elastic flagella \cite{Vogel2013,Son2013,Saggiorato2017,Kuhn2017}. Using this model, we can accurately assess the hydrodynamics of flagella near spheres, which is essential for understanding the mechanisms underlying bacterial motility in such fluid environments. Additionally, this model is suitable for studying bacterial motility on planar surfaces and within narrow channels \cite{Vizsnyiczai2020}. Moreover, it is well-suited for simulating the adaptive locomotion of artificial microswimmers \cite{Huang2019} and the targeted drug delivery capabilities of microrobots \cite{Xu2018}, offering valuable insights for the design and manufacturing of artificial microswimmer technologies.

This paper is organized as follows. In Sec. \uppercase\expandafter{\romannumeral2}, we describe the methodology for calculating the grand resistance matrix for a multi-sphere system and its inverse to obtain the grand mobility matrix. We demonstrate that the grand resistance matrix satisfies both non-overlapping and approximately no-slip boundary conditions. In Sec. \uppercase\expandafter{\romannumeral3}, we model a single rigid helical flagellum as a helical sequence of spheres, calculating the axial thrust and torque generated by the flagellum. Subsequently, we compare the results with existing laboratory measurements to validate the feasibility of the model and to identify an appropriate inter-sphere spacing. In Sec. \uppercase\expandafter{\romannumeral4}, we calculate the translational and rotational velocities of bacteria near a colloidal sphere using the RFT and TMM. This section evaluates the effects of hydrodynamic interactions among the bacterial cell bodies, flagella, and colloidal spheres on bacterial motility. Finally, we present our main conclusions in Sec. \uppercase\expandafter{\romannumeral5}.

\section{The resistance and mobility matrix}

As the characteristic size and speed of Escherichia coli bacteria in water are about $1$ \si{\mu m} and $10$ \si{\mu m/s}, respectively, the corresponding Reynolds number is approximately $10^{-4}\sim10^{-5}$. Therefore, bacterial flows are typically studied using the linear Stokes equations, which are given by \cite{Lauga2009,Happel2012,Kim2013}:
\begin{equation}
\begin{split}
&\mu\nabla^{2}\mathbf{u}-\nabla p+\mathbf{f}=0,\\
&\nabla\cdot\mathbf{u}=0
\label{eq:refname1}
\end{split}
\end{equation}
where $\mu$ is the dynamic viscosity, $\mathbf{u}$ is the fluid velocity, $p$ is the pressure, and $\mathbf{f}$ is the force applied to the fluid by the immersed body. Here, $\rho$, $U$, and $L$ are the fluid density, sphere velocity, and sphere dimension, respectively\cite{Kim2013}. Due to the linearity of the Stokes equations, the forces $\mathbf{F}$ and torques $\mathbf{T}$ that spheres exert on the fluid depend linearly on the translational velocities $\mathbf{U}$ and rotational velocities $\boldsymbol{\Omega}$ of the spheres. On account of the linearity of the Stokes equations, the resistance matrix that relates the forces and torques to the translational and rotational velocities of $N$ spheres is \cite{Durlofsky1989,Durlofsky1987,Brady1988}:
\begin{equation}
\left(\begin{matrix}\mathbf{F}\\ \mathbf{T}\end{matrix}\right)=\mathcal{R}\left(\begin{matrix}\mathbf{U}-\mathbf{U}^{\infty}\\ \boldsymbol{\Omega}-\boldsymbol{\Omega}^{\infty}\end{matrix}\right).
\label{eq:refname2}
\end{equation}
where $\mathbf{U}^{\infty}$ and $\boldsymbol{\Omega}^{\infty}$ are the ambient flow fields, and $\mathbf{U}-\mathbf{U}^{\infty}$ and $\boldsymbol{\Omega}-\boldsymbol{\Omega}^{\infty}$ are velocities of dimension $3N$ of all $N$ spheres relative to the ambient flow. The inverse problem is to calculate the velocities of spheres given their forces and torques:
\begin{equation}
\left(\begin{matrix}\mathbf{U}\\ \boldsymbol{\Omega}\end{matrix}\right)=\left(\begin{matrix}\mathbf{U}^{\infty}\\ \boldsymbol{\Omega}^{\infty} \end{matrix}\right)+\mathcal{M}\left(\begin{matrix}\mathbf{F}\\ \mathbf{T}\end{matrix}\right).
\label{eq:refname3}
\end{equation}

\begin{figure}
\centering
   \includegraphics[width=0.5\textwidth]{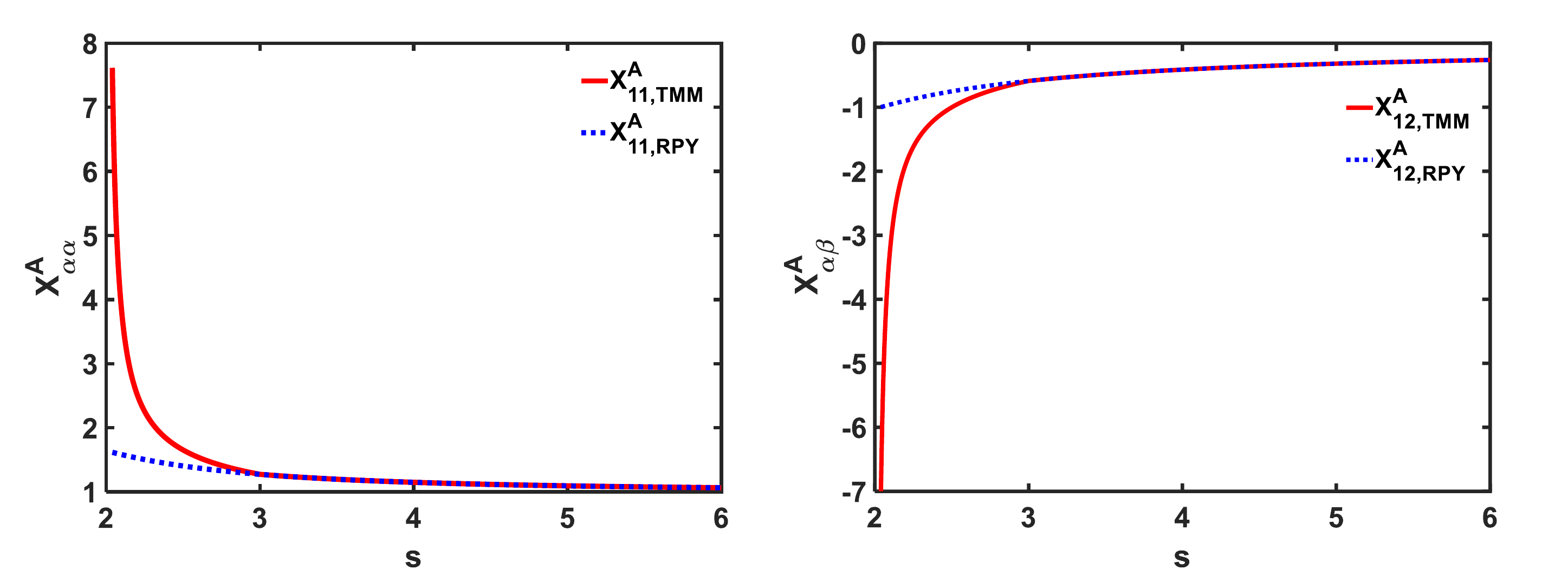}
    \caption{Comparison of longitudinal resistance functions as functions of the variable $s$, obtained using the TMM and the RPY tensor, respectively. The resistance function $X_{\alpha\alpha}^{A}$ couples the longitudinal force of one sphere with its own longitudinal translational velocity, while $X_{\alpha\beta}^{A}$ couples the longitudinal force of one sphere with the longitudinal translational velocity of the other sphere.}\label{fig:fig1}
\end{figure}

\begin{figure}
\centering
   \includegraphics[width=0.5\textwidth]{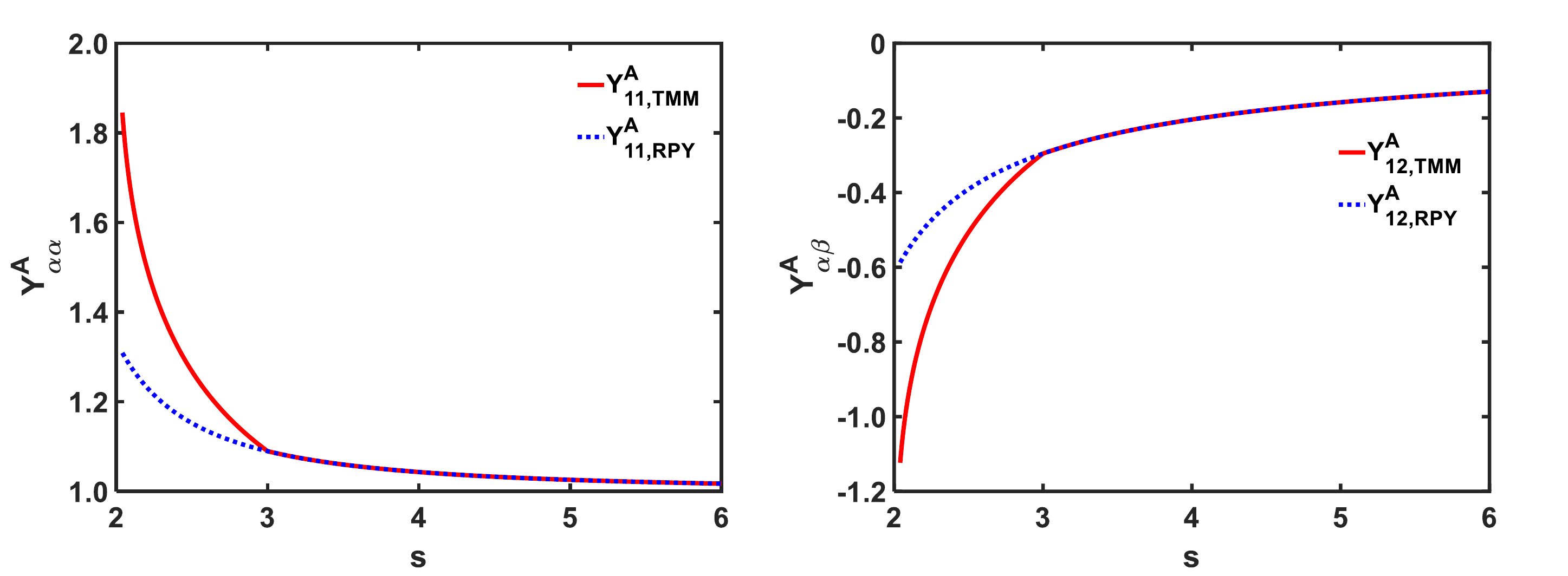}
    \caption{Comparison of transverse resistance functions as functions of the variable $s$, obtained using the TMM and the RPY tensor, respectively. The resistance function $Y_{\alpha\alpha}^{A}$ couples the transverse force of one sphere with its own transverse translational velocity, while $Y_{\alpha\beta}^{A}$ couples the transverse force of one sphere with the transverse translational velocity of the other sphere.}\label{fig:fig2}
\end{figure}

\begin{figure}
\centering
   \includegraphics[width=0.5\textwidth]{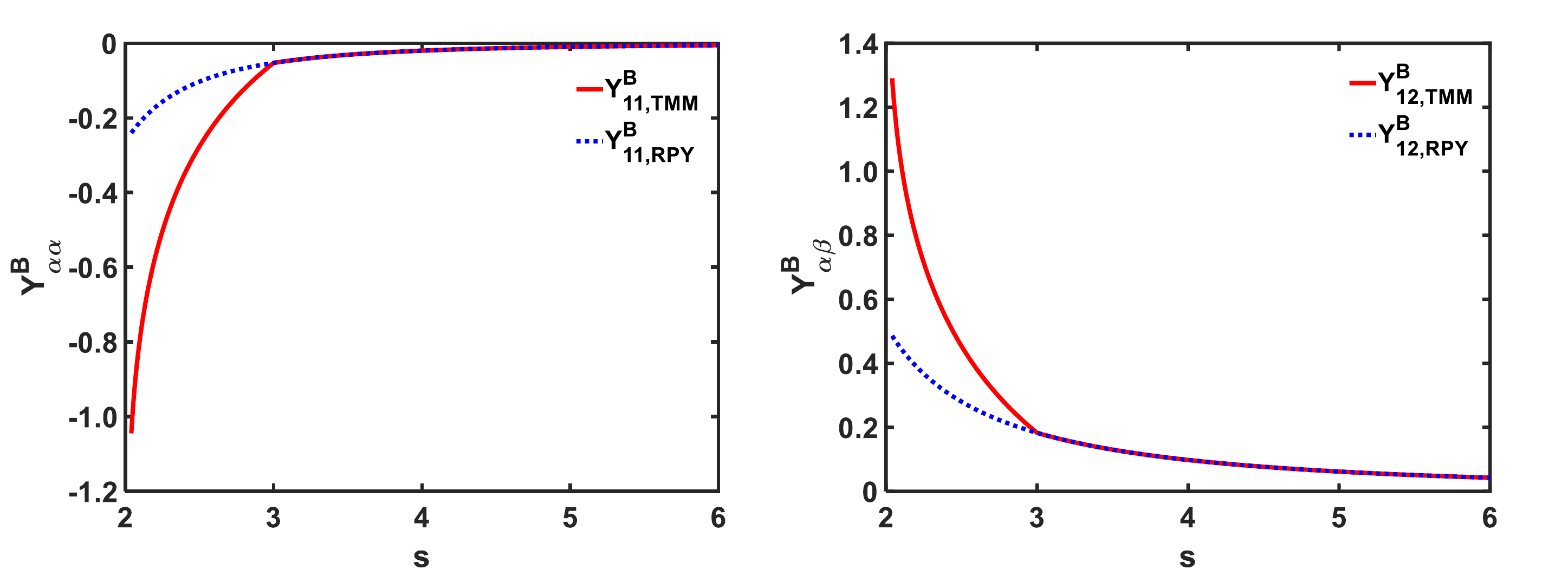}
    \caption{Comparison of the resistance functions $Y_{\alpha\alpha}^{B}$ and $Y_{\alpha\beta}^{B}$ as functions of the variable $s$, obtained using the TMM and the RPY tensor, respectively. The resistance function $Y_{\alpha\alpha}^{B}$ couples the torque of one sphere with its own translational velocity, while the resistance function $Y_{\alpha\beta}^{B}$ couples the torque of one sphere with the translational velocity of the other sphere.}\label{fig:fig3}
\end{figure}

\begin{figure}
\centering
   \includegraphics[width=0.5\textwidth]{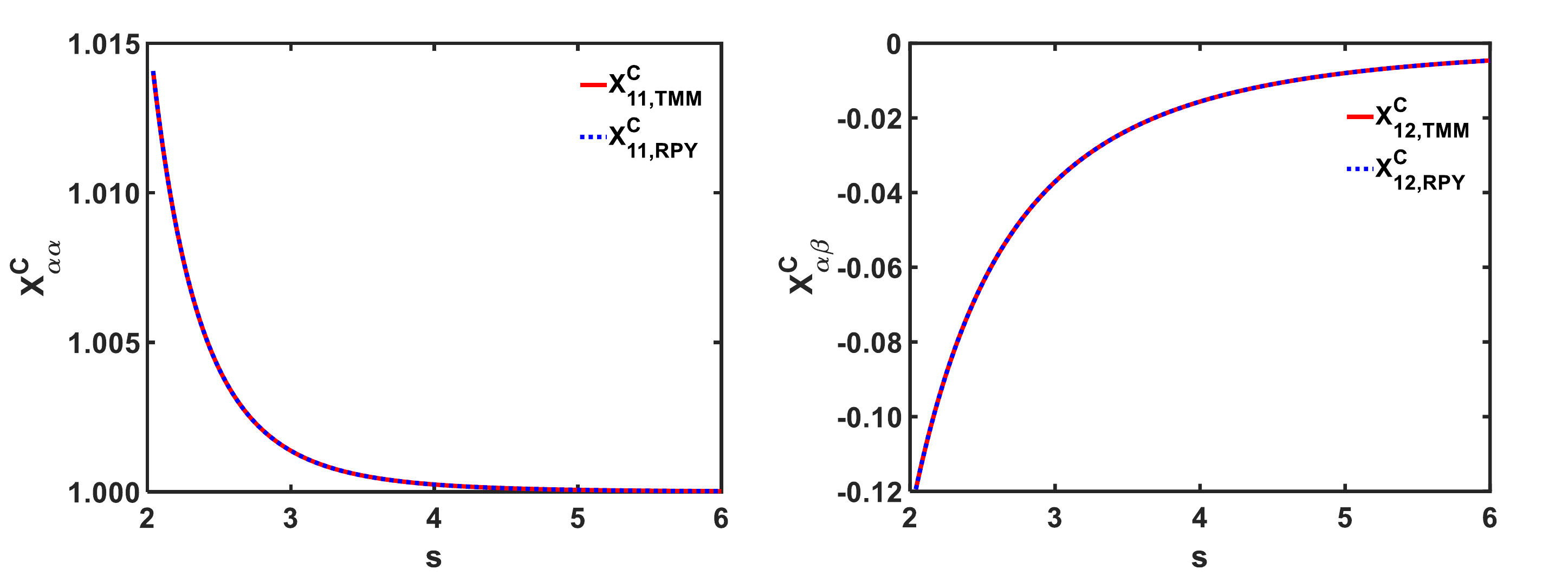}
    \caption{Comparison of longitudinal resistance functions as functions of the variable $s$, obtained using the TMM and the RPY tensor, respectively. The resistance function $X_{\alpha\alpha}^{C}$ couples the longitudinal torque of one sphere with its own longitudinal rotational velocity, while $X_{\alpha\beta}^{C}$ couples the longitudinal torque of one sphere with the longitudinal rotational velocity of the other sphere.}\label{fig:fig4}
\end{figure}

\begin{figure}
\centering
   \includegraphics[width=0.5\textwidth]{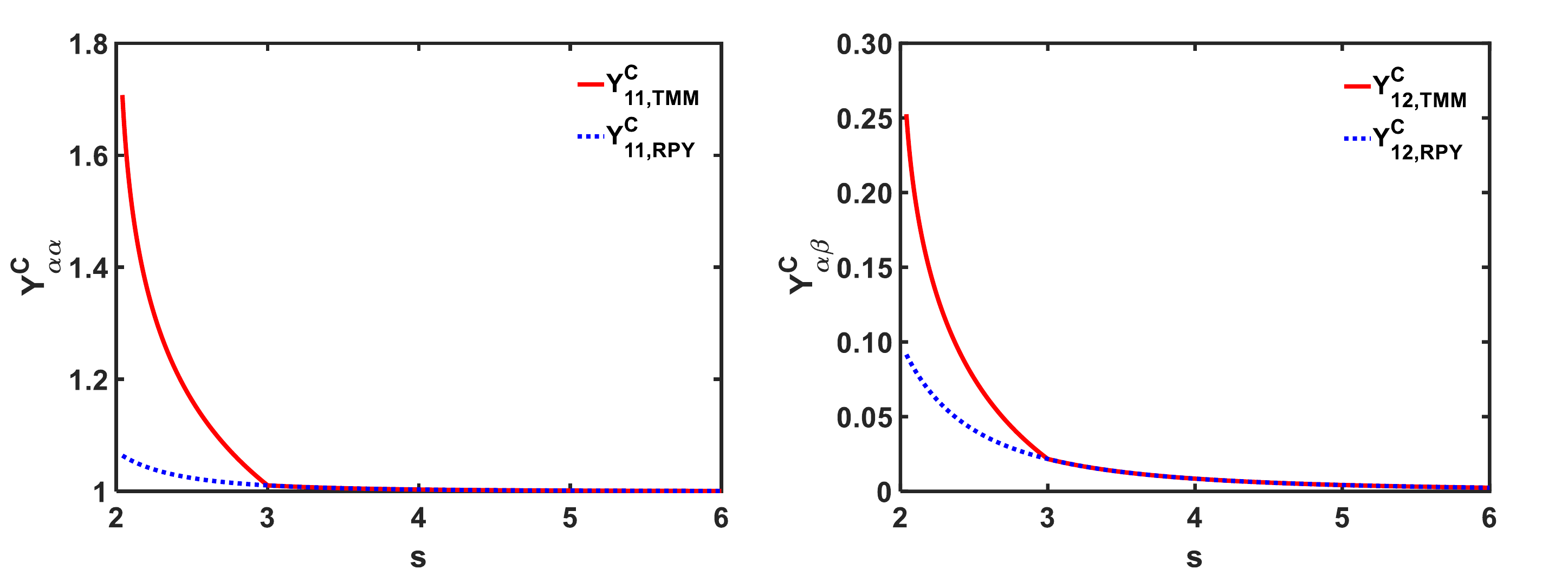}
    \caption{Comparison of transverse resistance functions as functions of the variable $s$, obtained using the TMM and the RPY tensor, respectively. The resistance function $Y_{\alpha\alpha}^{C}$ couples the transverse torque of one sphere with its own transverse rotational velocity, while $Y_{\alpha\beta}^{C}$ couples the transverse torque of one sphere with the transverse rotational velocity of the other sphere.}\label{fig:fig5}
\end{figure}

Consider a set of $N$ rigid spheres of arbitrary size immersed in a viscous fluid. Using the TMM, the grand resistance matrix $\mathcal{R}$, which contains both near-field lubrication effects and far-field hydrodynamic interactions, can be constructed as
\begin{equation}
\begin{split}
\mathcal{R}=(\mathcal{M}^{\infty})^{-1}+\mathcal{R}_{2B,lub}.
\label{eq:refname4}
\end{split}
\end{equation}
where the subscript 'lub' denotes 'lubrication'. The grand resistance matrix encompasses both far-field interactions, achieved through the inversion of $\mathcal{M}^{\infty}$, and pairwise lubrication interactions $\mathcal{R}_{2B,lub}$. We define non-dimensional variables $s=2r/(a_{\alpha}+a_{\beta})$, $\xi=s-2$, and $\lambda=a_{\beta}/a_{\alpha}$ (the subscripts $\alpha$ and $\beta$ indicate the labels of the spheres, which take all values from $1$ to $N$), where $r$ is the center-to-center distance between the spheres and $a_{\alpha}$ is the radius of sphere $\alpha$. The far-field mobility matrix for multi-sphere system, $\mathcal{M}^{\infty}$, is presented as
\begin{equation}
\begin{split}
\mathcal{M}^{\infty}=\left(\begin{matrix} \mathbf{M}^{tt}_{\alpha\beta} & \mathbf{M}^{tr}_{\alpha\beta} \\ \mathbf{M}^{rt}_{\alpha\beta} & \mathbf{M}^{rr}_{\alpha\beta} \end{matrix}\right).
\label{eq:refname5}
\end{split}
\end{equation}
$\mathcal{M}^{\infty}$ is a far-field approximation to the interaction between spheres and includes terms up to $O(s^{-3})$, which is exactly the same as that of the RPY tensor \cite{Wajnryb2013}. The detailed elements of $\mathcal{M}^{\infty}$ are provided explicitly in Appendix A, following the notation of Jeffrey and Onishi \cite{Jeffrey1984}. The inverse of $\mathcal{M}^{\infty}$ includes interactions among multiple spheres. The pairwise near-field lubrication resistance matrix is denoted as $\mathcal{R}_{2B,lub}$ and is constructed as follows
\begin{equation}
\begin{split}
\mathcal{R}_{2B,lub}=\left(\begin{matrix} \mathbf{R}^{tt}_{\alpha\beta} & \mathbf{R}^{tr}_{\alpha\beta} \\ \mathbf{R}^{rt}_{\alpha\beta} & \mathbf{R}^{rr}_{\alpha\beta} \end{matrix}\right).
\label{eq:refname6}
\end{split}
\end{equation}
for $\xi<\min\{\lambda,1/\lambda\}$ and is a zero matrix for $\xi\geqslant\min\{\lambda,1/\lambda\}$. The segmentation point between the near- and far-field can be defined as $s=2+\min\{\lambda,1/\lambda\}$ or a smaller value. $\mathcal{R}_{2B,lub}$ is a continuous function of the variable $s$ and is equal to zero when $s\geqslant2+\min\{\lambda,1/\lambda\}$ to ensure the continuity of the grand resistance matrix. This procedure captures both near- and far-field interactions and reduces computational costs. The vector $\mathbf{e}_{\alpha\beta}=\left(\mathbf{x}_{\alpha}-\mathbf{x}_{\beta}\right)/\left|\mathbf{x}_{\alpha}-\mathbf{x}_{\beta}\right|$ is the unit vector along the line of centers. These submatrices $\mathbf{R}^{tt}_{\alpha\beta}$, $\mathbf{R}^{tr}_{\beta\alpha}$, $\mathbf{R}^{rt}_{\alpha\beta}$, $\mathbf{R}^{rr}_{\alpha\beta}$, $\mathbf{M}^{tt}_{\alpha\beta}$, $\mathbf{M}^{tr}_{\alpha\beta}$, $\mathbf{M}^{rt}_{\alpha\beta}$, and $\mathbf{M}^{rr}_{\alpha\beta}$ are second-rank tensors. We define the non-dimensional matrices as
\begin{equation}
\begin{split}
&\widehat{\mathbf{R}}_{\alpha\beta}^{tt}=\frac{\mathbf{R}_{\alpha\beta}^{tt}}{3\pi\mu(a_{\alpha}+a_{\beta})},\quad
\widehat{\mathbf{M}}_{\alpha\beta}^{tt}=3\pi\mu(a_{\alpha}+a_{\beta})\mathbf{M}_{\alpha\beta}^{tt},\\
&\widehat{\mathbf{R}}_{\alpha\beta}^{rt}=\frac{\mathbf{R}_{\alpha\beta}^{rt}}{\pi\mu(a_{\alpha}+a_{\beta})^{2}},\quad
\widehat{\mathbf{M}}_{\alpha\beta}^{rt}=\pi\mu(a_{\alpha}+a_{\beta})^{2}\mathbf{M}_{\alpha\beta}^{rt},\\
&\widehat{\mathbf{R}}_{\alpha\beta}^{rr}=\frac{\mathbf{R}_{\alpha\beta}^{rr}}{\pi\mu(a_{\alpha}+a_{\beta})^{3}},\quad
\widehat{\mathbf{M}}_{\alpha\beta}^{rr}=\pi\mu(a_{\alpha}+a_{\beta})^{3}\mathbf{M}_{\alpha\beta}^{rr}.
\label{eq:refname7}
\end{split}
\end{equation}
Then the non-dimensional submatrices can be expressed as follows, respectively
\begin{equation}
\begin{split}
&\widehat{\mathbf{R}}_{\alpha\beta}^{tt}=X_{\alpha\beta}^{a}\mathbf{e}_{\alpha\beta}\otimes\mathbf{e}_{\alpha\beta}
+Y_{\alpha\beta}^{a}(\mathbb{I}-\mathbf{e}_{\alpha\beta}\otimes\mathbf{e}_{\alpha\beta}),\\
&\widehat{\mathbf{R}}_{\alpha\beta}^{rt}=\left[\widehat{\mathbf{R}}_{\beta\alpha}^{tr}\right]^{T}=Y_{\alpha\beta}^{b}(\mathbf{e}_{\alpha\beta}\times),\\
&\widehat{\mathbf{R}}_{\alpha\beta}^{rr}=X_{\alpha\beta}^{c}\mathbf{e}_{\alpha\beta}\otimes\mathbf{e}_{\alpha\beta}
+Y_{\alpha\beta}^{c}(\mathbb{I}-\mathbf{e}_{\alpha\beta}\otimes\mathbf{e}_{\alpha\beta}).
\label{eq:refname8}
\end{split}
\end{equation}
and
\begin{equation}
\begin{split}
&\widehat{\mathbf{M}}_{\alpha\beta}^{tt}=x_{\alpha\beta}^{a}\mathbf{e}_{\alpha\beta}\otimes\mathbf{e}_{\alpha\beta}
+y_{\alpha\beta}^{a}(\mathbb{I}-\mathbf{e}_{\alpha\beta}\otimes\mathbf{e}_{\alpha\beta}),\\
&\widehat{\mathbf{M}}_{\alpha\beta}^{rt}=\left[\widehat{\mathbf{M}}_{\beta\alpha}^{tr}\right]^{T}=y_{\alpha\beta}^{b}(\mathbf{e}_{\alpha\beta}\times),\\
&\widehat{\mathbf{M}}_{\alpha\beta}^{rr}=x_{\alpha\beta}^{c}\mathbf{e}_{\alpha\beta}\otimes\mathbf{e}_{\alpha\beta}
+y_{\alpha\beta}^{c}(\mathbb{I}-\mathbf{e}_{\alpha\beta}\otimes\mathbf{e}_{\alpha\beta}).
\label{eq:refname9}
\end{split}
\end{equation}
where $\mathbb{I}$ is a $3\times3$ identity matrix, $\otimes$ is the dyadic product, and $(\mathbf{e}\times)$ is a $3\times3$ antisymmetric matrix defined such that $(\mathbf{e}\times)\mathbf{v}=\mathbf{e}\times\mathbf{v}$ for any vector $\mathbf{v}$. The elements of the matrix obey a series of symmetry conditions, as detailed in Jeffrey and Onishi \cite{Jeffrey1984}. The specific expressions of resistance and mobility functions are detailed in Appendix A. The grand resistance matrix $\mathcal{R}$, expressed in terms of its individual elements is
\begin{equation}
\begin{split}
\mathcal{R}=\left(\begin{matrix} \mathcal{R}^{tt}_{\alpha\beta} & \mathcal{R}^{tr}_{\alpha\beta} \\ \mathcal{R}^{rt}_{\alpha\beta} & \mathcal{R}^{rr}_{\alpha\beta} \end{matrix}\right).
\label{eq:refname10}
\end{split}
\end{equation}
where the non-dimensional submatrices of $\widehat{\mathcal{R}}$ are expressed as:
\begin{equation}
\begin{split}
&\widehat{\mathcal{R}}_{\alpha\beta}^{tt}=X_{\alpha\beta}^{A}\mathbf{e}_{\alpha\beta}\otimes\mathbf{e}_{\alpha\beta}
+Y_{\alpha\beta}^{A}(\mathbb{I}-\mathbf{e}_{\alpha\beta}\otimes\mathbf{e}_{\alpha\beta}),\\
&\widehat{\mathcal{R}}_{\alpha\beta}^{rt}=\left[\widehat{\mathcal{R}}_{\beta\alpha}^{tr}\right]^{T}=Y_{\alpha\beta}^{B}(\mathbf{e}_{\alpha\beta}\times),\\
&\widehat{\mathcal{R}}_{\alpha\beta}^{rr}=X_{\alpha\beta}^{C}\mathbf{e}_{\alpha\beta}\otimes\mathbf{e}_{\alpha\beta}
+Y_{\alpha\beta}^{C}(\mathbb{I}-\mathbf{e}_{\alpha\beta}\otimes\mathbf{e}_{\alpha\beta}).
\label{eq:refname11}
\end{split}
\end{equation}
The scalar grand resistance functions $X^{A}$, $Y^{A}$, $Y^{B}$, $X^{C}$, and $Y^{C}$ are plotted versus the reduced inter-sphere separation distance, $s=2r/(a_{\alpha}+a_{\beta})$, in Figs.~\ref{fig:fig1}-\ref{fig:fig5}. The grand mobility matrix is defined as the inverse of the grand resistance matrix
\begin{equation}
\begin{split}
\mathcal{M}=\mathcal{R}^{-1}.
\label{eq:refname12}
\end{split}
\end{equation}

The grand mobility matrix $\mathcal{M}$ obtained by this method is convergent at any reduced separation distance $s$. To account for near-field lubrication effects, the term $\xi^{-1}$ is introduced in the longitudinal resistance function $X^{A}_{\alpha\alpha}$ to avoid overlapping spheres in numerical simulations. We examine the results for the resistance functions and compare them with those obtained from the RPY tensor where applicable. Several resistance functions of two equal spheres are plotted as functions of the reduced separation distance $s$ in Figs.~\ref{fig:fig1}-\ref{fig:fig5}. These figures are generated in the region $2.05\leqslant s\leqslant6.0$.

The longitudinal components $X^{A}$ of the resistance matrix $\mathcal{R}^{tt}$ are plotted in Fig.~\ref{fig:fig1}. These components couple the longitudinal forces and the longitudinal translational velocities. The scalar resistance function $X^{A}_{\alpha\beta}$ indicates that the translation of one sphere in the longitudinal direction exerts a force in the longitudinal direction on the other spheres. When the spheres come into contact with each other, that is, $s=2$, the scalar resistance functions $X^{A}$ tend to infinity, which automatically prevents the overlapping of the two spheres. These components include near-field lubrication effects, which differ significantly from those of the RPY tensor. Physically, when the other spheres are far away, a sphere can move freely, with the self-interaction component $X^{A}_{\alpha\alpha}\to1$ and the pair interaction component $X^{A}_{\alpha\beta}\to0$. The longitudinal components $X^{A}$ reveal that for a system of two spheres, when equal but opposite forces are applied in the longitudinal direction, the velocities at which the spheres approach or move away from each other decreases as the distance between them decreases. This process continues until they are in complete contact, at which point the longitudinal components $X^{A}$ become infinite.

The transverse components $Y^{A}$ of the resistance matrix $\mathcal{R}^{tt}$ are plotted in Fig.~\ref{fig:fig2}. These components couple the transverse forces and the transverse translational velocities. $Y^{A}_{\alpha\beta}$ indicates that the translation of one sphere in the transverse direction exerts a force in the transverse direction on the other spheres. When a pair of spheres is widely separated, the self-interaction component $Y^{A}_{\alpha\alpha}\to1$ and the pair interaction component $Y^{A}_{\alpha\beta}\to0$. The transverse components $Y^{A}$ reveal that in a system of two spheres, when one sphere is stationary and the other sphere translates transversely, the moving sphere experiences a resistance from the stationary sphere, and this resistance increases as the distance between the spheres decreases.

The scalar resistance functions $Y^{B}$ in Fig.~\ref{fig:fig3} illustrate the coupling between forces and rotational velocities, as well as between torques and translational velocities. When a pair of spheres is widely separated, a hydrodynamic force or torque acting on one sphere will not induce any rotation or translation of the other sphere. This means that the components $Y^{B}_{\alpha\alpha}\to 0$ and $Y^{B}_{\alpha\beta}\to0$ when $s\to\infty$. However, as one sphere approaches another, the drag force or torque on one sphere will induce rotation or translation of the other. Specifically, the rotation of one sphere exerts a force on the other sphere, while the translation of one sphere exerts a torque on the other sphere. 

The longitudinal components $X^{C}$ of the resistance matrix $\mathcal{R}^{rr}$ are plotted in Fig.~\ref{fig:fig4}, which couple longitudinal torques and longitudinal rotational velocities. Fig.~\ref{fig:fig4} shows that the scalar resistance functions $X^{C}$ obtained using the TMM are identical to those obtained using the RPY tensor. The self-interaction component $X^{C}_{\alpha\alpha}\to1$ and the pair interaction component $X^{C}_{\alpha\beta}\to0$ when a pair of spheres is far apart. The transverse components $Y^{C}$ of the resistance matrix $\mathcal{R}^{rr}$ are plotted in Fig.~\ref{fig:fig5}, which provide the coupling between torques and rotational velocities perpendicular to the line of centers. The components $X^{C}$ and $Y^{C}$ indicate that the rotation of one sphere exerts torques on its neighboring spheres.

\begin{figure}
\centering
   \includegraphics[width=0.5\textwidth]{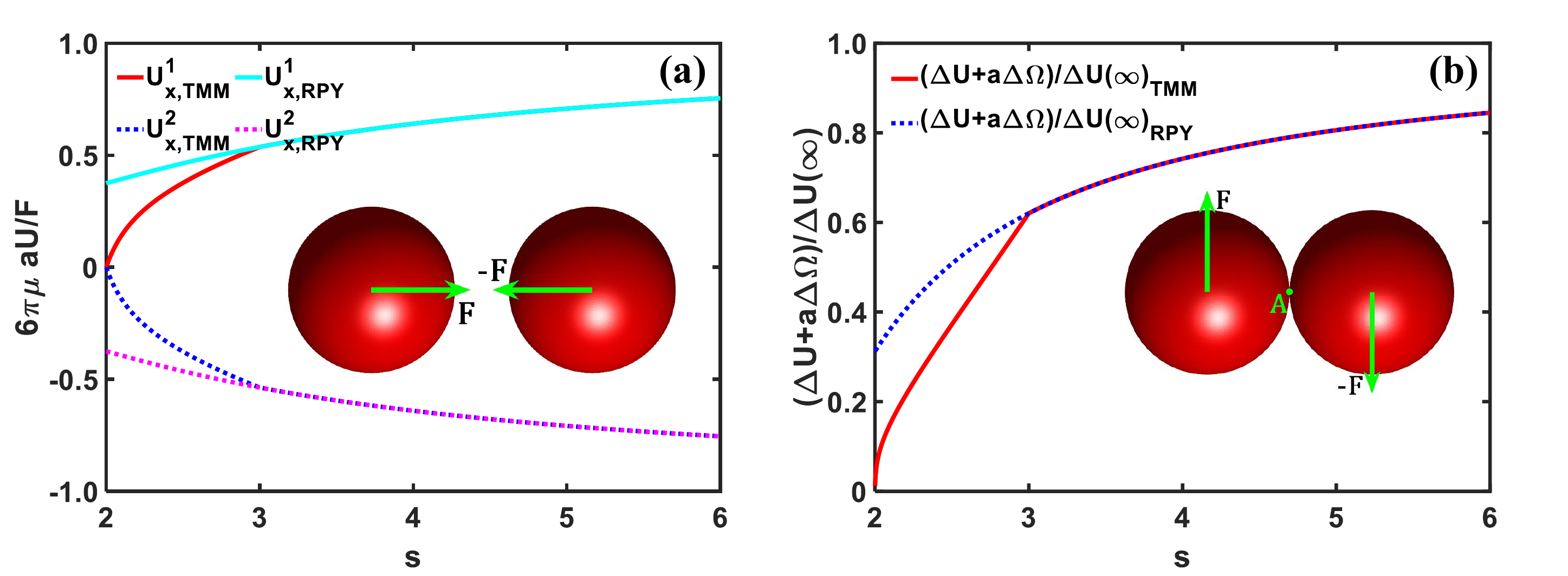}
    \caption{(a) The translational velocities of two equal spheres subject to forces $\mathbf{F}$ and $-\mathbf{F}$ along the $x$-axis as functions of the reduced separation distance $s$. (b) The velocity difference of two equal spheres subject to forces $\mathbf{F}$ and $-\mathbf{F}$ along the $y$-axis as a function of the reduced separation distance $s$. The results are obtained using the TMM and the RPY tensor, respectively.}\label{fig:fig6}
\end{figure}

As an example, in Fig.~\ref{fig:fig6}(a), we compare the translational velocities of two equal spheres subjected to forces $\mathbf{F}$ and $-\mathbf{F}$, calculated using the TMM and the RPY tensor, respectively. Instead of moving very slowly toward each other, the two spheres would physically touch and overlap when using the RPY tensor. In contrast, using the TMM, we find that the translational velocities gradually approach zero as $s$ approaches $2.0$. This result differs significantly from that obtained using the RPY tensor and reveals the importance of near-field lubrication effects. The grand resistance matrix calculated using the TMM automatically prevents sphere overlap, eliminating the need to introduce a repulsive potential \cite{Loi2011,Schlick1992,Jendrejack2003,Jendrejack2004,Schroeder2004,Ando2013}.

The prediction for the no-slip boundary condition is $\left[(U_{y}^{1}-U_{y}^{2})+a(\Omega_{z}^{1}+\Omega_{z}^{2})\right]/\left(U_{y}^{1}(\infty)-U_{y}^{2}(\infty)\right)=0$ at $s=2$, where $U_{y}^{1}(\infty)=F/6\pi\mu a$ and $U_{y}^{2}(\infty)=-F/6\pi\mu a$, with $a$ being the radius of the sphere, $\mu$ is the dynamic viscosity, and $\mathbf{U}$ and $\boldsymbol{\Omega}$ representing the translational and rotational velocities of the centers of the spheres, respectively. To demonstrate that the grand resistance matrix calculated using the TMM approximately satisfies the no-slip boundary condition, two equal spheres are lined up along the $x$-axis, and the forces $\mathbf{F}$ and $-\mathbf{F}$ are applied to the two spheres along the $y$-axis. The numerical simulation matches our prediction at a separation distance of $s\to 2$, as shown in Fig.~\ref{fig:fig6}(b). This indicates that the no-slip boundary condition has been approximately achieved at the closest point 'A', where the transverse translation vanishes but rotation remains possible. The slight deviation of the result from zero as $s\to2$ is due to neglecting higher-order terms in the calculation of the matrix $\mathcal{R}$, these terms increase computational cost but have little impact on the results. The results obtained using the RPY tensor differ significantly from our prediction.

\section{Propulsion of a helical flagellum}
\hangafter=-1\hangindent=19pt\noindent

\begin{figure}
\centering
   \includegraphics[width=0.45\textwidth]{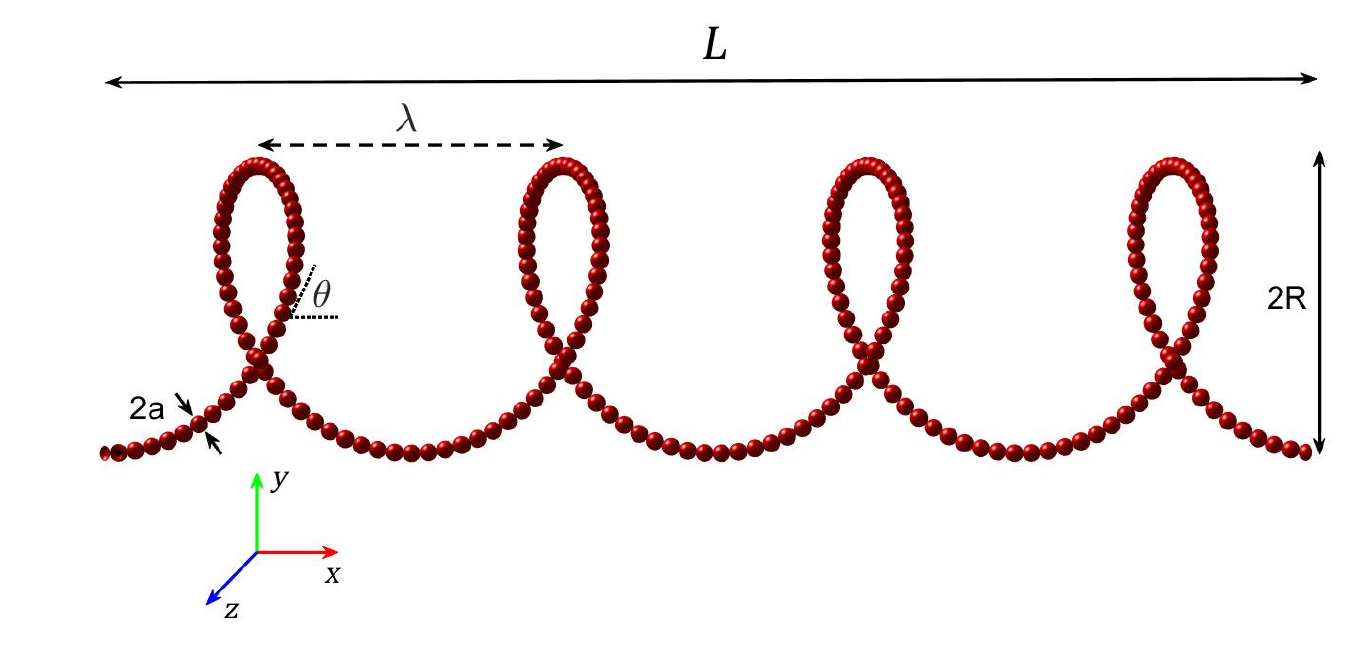}
    \caption{Schematic of a helical flagellum model. The helix radius of the flagellum is $R$, the helix pitch is $\lambda$, the axial length is $L$, the filament radius is $a$, the contour length is $\Lambda=L/\cos\theta$, and the pitch angle is $\theta$, where $\tan\theta=2\pi R/\lambda$.}\label{fig:fig7}
\end{figure}

The primary function of flagella in bacteria is to facilitate motility. In our model, we represent the bacterial flagellum as a helical sequence of spheres to determine the axial thrust and torque. Due to the isotropic nature of the spheres and the assumption of neglecting hydrodynamic interactions among them, a rotating, non-translating flagellum would not produce thrust for bacteria in a viscous fluid. Therefore, it is essential to incorporate the hydrodynamic interactions among the spheres in our analysis. In this section, we examine the relationship between the axial thrust and torque generated by a non-translating rigid helical flagellum rotating in a viscous fluid and its pitch, denoted as $\lambda$. A left-handed helical flagellum can be modeled as a rigid rotating helix characterized by several parameters: helical radius $R$, pitch $\lambda$, axial length $L$, filament radius $a$, and pitch angle $\theta$. The contour length is given by $\Lambda=L/\cos\theta$, where $\tan\theta=2\pi R/\lambda$, as illustrated in Fig.~\ref{fig:fig7}. The centerline of the flagellum is defined as follows: 
\begin{equation}
\begin{split}
\mathbf{r}(s)=[x,\sin(kx+\Omega t),\cos(kx+\Omega t)],\quad x\in[0,L].
\label{eq:refname13}
\end{split}
\end{equation}
where $x=l\cos\theta$, $l$ is the arc length of the flagellar centerline, $\Omega$ is the rotation rate, and $k=2\pi/\lambda$ is the wave number. At low Reynolds number, the translational velocity $\mathbf{U}$ and rotational velocity $\boldsymbol{\Omega}$ are linearly related to the force $\mathbf{F}$ and torque $\mathbf{T}$ exerted on the fluid by the flagellum, as described by Eq.~\ref{eq:refname2}. The resistance matrix in this equation depends only on the geometry of the flagellum. We calculate the resistance matrix of the flagellum using the TMM and LSBT, respectively. LSBT models the flagellum as a series of Stokeslets and doublets distributed along its centerline, which is parameterized by the arc length $l$. For the fluid loading analysis, LSBT relates the local velocity $\mathbf{u}(s)$ to the force per unit length $\mathbf{f}(s)$ at each point along the flagellar centerline \cite{Rodenborn2013,Jawed2015}:
\begin{equation}
\begin{split}
-\mathbf{u}(s)=\frac{\mathbf{f}_{\perp}(s)}{4\pi\mu}+\int_{|\mathbf{r}(s',s)|>\delta}\mathbf{f}(s')\cdot \mathbb{J}(\mathbf{r}(s'))ds'.
\label{eq:refname14}
\end{split}
\end{equation}
where $\mathbf{f}_{\perp}(s)=\mathbf{f}(s)\cdot[\mathbb{I}-\mathbf{t}(s)\otimes\mathbf{t}(s)]$ is the component of $\mathbf{f}$ in the plane perpendicular to the tangent $\mathbf{t}(s)$, $\mathbf{r}(s',s)$ is the position vector from $s'$ to $s$, $\delta=a\sqrt{e}/2$ is the natural cutoff length and $e$ is the natural constant. The minus sign indicates the force exerted on the fluid by the flagellum. $\mathbb{J}$ is the Oseen tensor given by:
\begin{equation}
\begin{split}
\mathbb{J}(\mathbf{r})\equiv\frac{1}{8\pi\mu}\left(\frac{\mathbb{I}}{|\mathbf{r}|}+\frac{\mathbf{r}\otimes\mathbf{r}}{|\mathbf{r}|^{3}}\right).
\label{eq:refname15}
\end{split}
\end{equation}

\begin{figure}
\centering
   \includegraphics[width=0.45\textwidth]{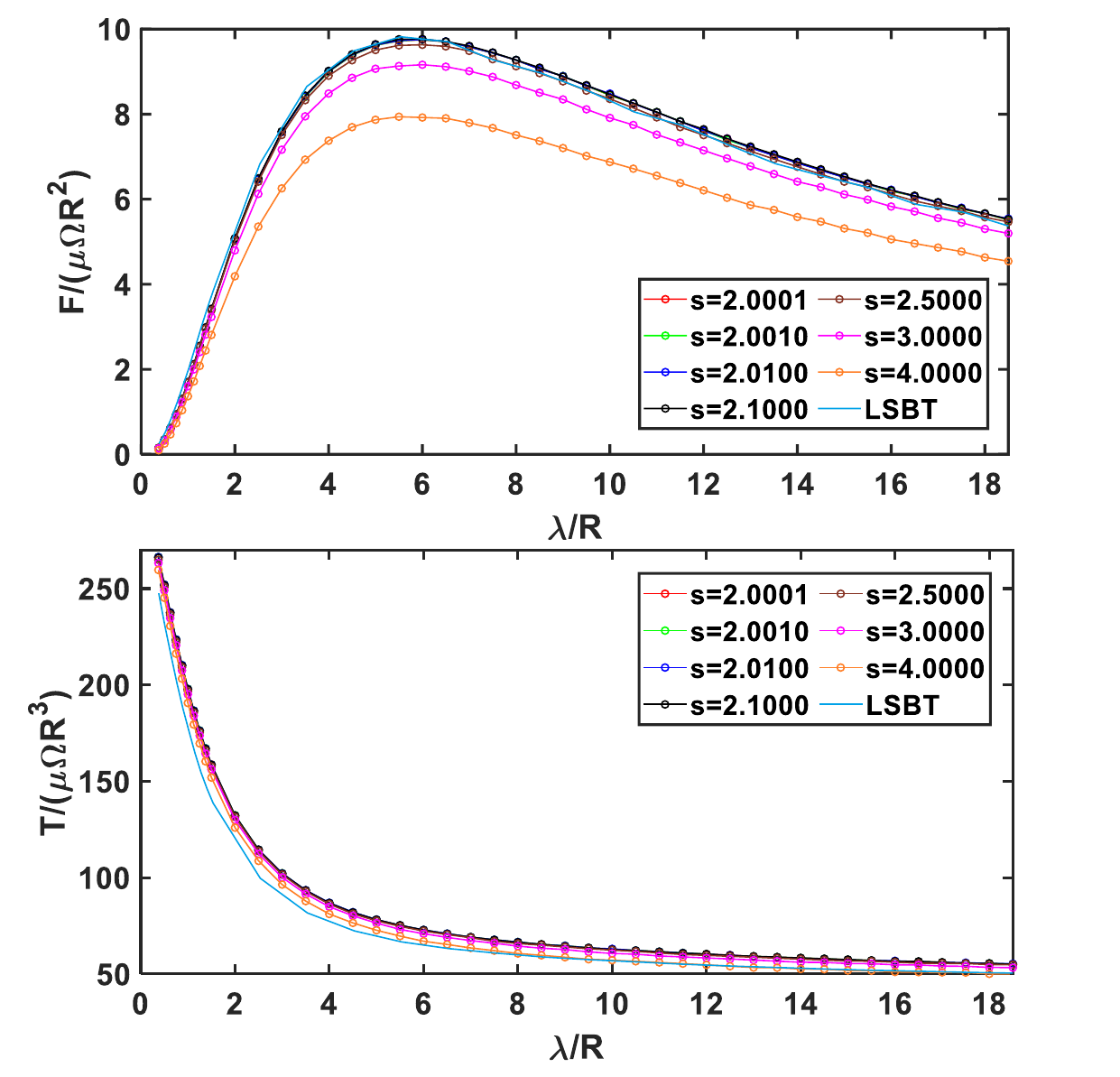}
    \caption{Non-dimensional thrust and torque for a helical flagellum are shown as functions of pitch $\lambda$ with axial length $L=20R$ and filament radius $a=R/16$ for various values of $s$. We indicate the results of LSBT with cyan solid lines.}\label{fig:fig8}
\end{figure}

The non-dimensional axial thrust $F/(\mu\Omega R^{2})$ and torque $T/(\mu\Omega R^{3})$ for a rotating, non-translating flagellum are illustrated in Fig.~\ref{fig:fig8}. These results are obtained using the TMM for varying sphere spacing in the range of $2.0001\leqslant s\leqslant4.0$. The work of Rodenborn et al. demonstrates that the calculated results of the LSBT are consistent with the laboratory measurements \cite{Rodenborn2013}. Our findings indicate that a sphere spacing of $2.0001\leqslant s\leqslant2.5$ is optimal, which aligns with the results obtained from the LSBT. This concordance supports our approach of modeling the flagellum as a helical sequence of spheres and employing the TMM to effectively compute the resistance matrix of the microswimmer system. Notably, this model is applicable not only to rigid flagella but also to the nonlinear dynamics of elastic filaments \cite{Loi2011,Vogel2013,Son2013,Saggiorato2017,Kuhn2017,Schlick1992,Jendrejack2003,Jendrejack2004,Schroeder2004,Ando2013,Jawed2015}.

\section{Microswimmer swimming near a colloidal sphere}
\hangafter=-1\hangindent=19pt\noindent

\begin{figure}
\centering
   \includegraphics[width=0.40\textwidth]{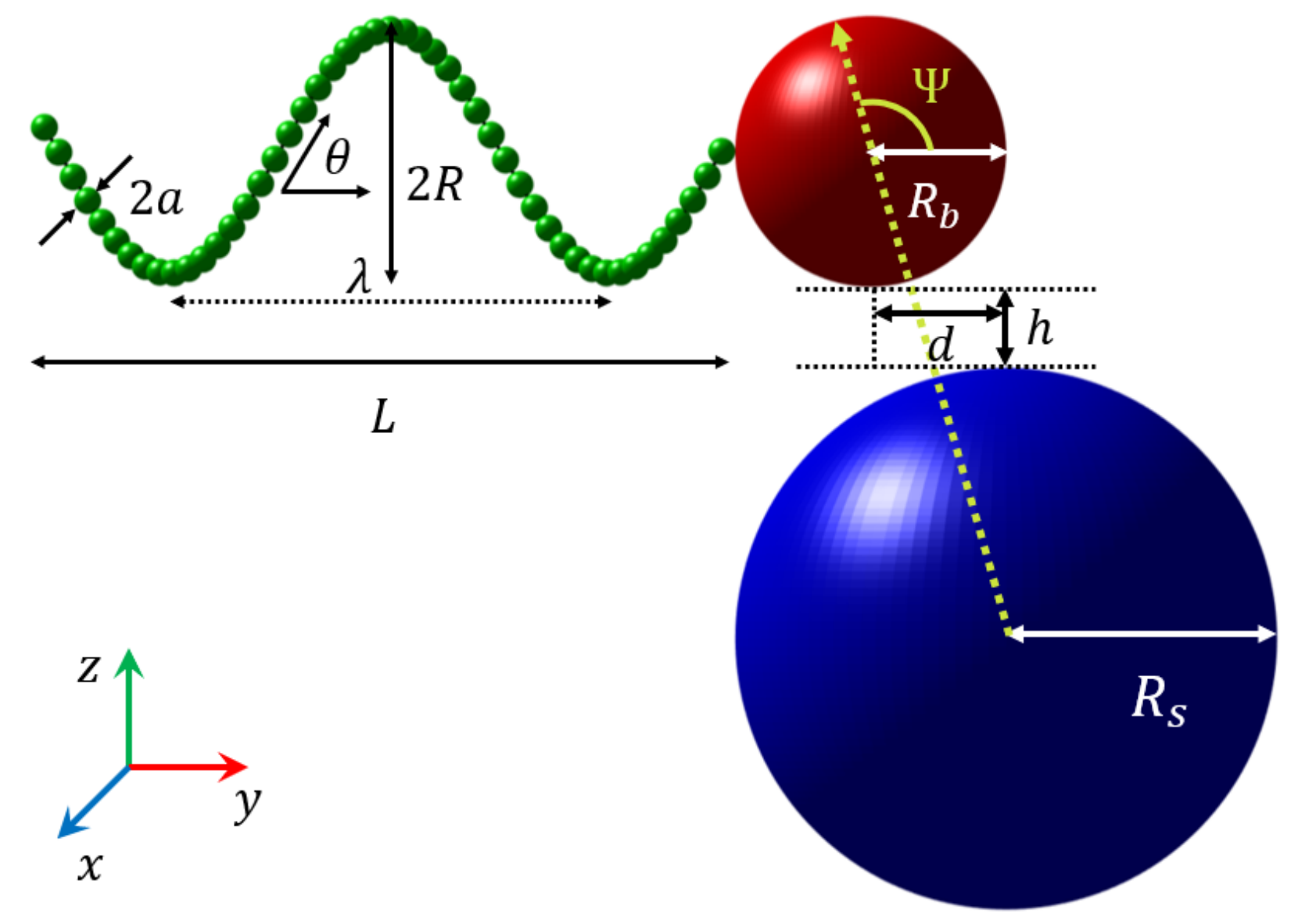}
    \caption{Illustration of a microswimmer model near a colloidal sphere in the fixed frame $\left\{x,y,z\right\}$. The flagellum axis is along the $y$-axis direction. The radii of the cell body and the colloidal sphere are $R_{b}$ and $R_{s}$, respectively. The gap between the cell body and the colloidal sphere along the $z$-axis is $h$, and the distance along the $y$-axis is $d$. The inclination angle of flagellar axis relative to the line of centers of the cell body and the colloidal sphere is $\Psi$.}\label{fig:fig9}
\end{figure}

The monotrichous microswimmer model depicted in Fig.~\ref{fig:fig9} serves as the bacterial model for this study and is positioned near a colloidal sphere. The helical, rigid flagellum is attached to the spherical cell body. The geometric parameters of the microswimmer model are detailed in Fig.~\ref{fig:fig9}. The radii of the cell body and the colloidal sphere are $R_{b}=1$ \si{\mu m} and $R_{s}=50$ \si{\mu m}, respectively. The flagellum is modeled as a rigid, rotating helix with helical radius $R=0.90$ \si{\mu m}, axial length $L=6.0$ \si{\mu m}, pitch angle $\theta=\pi/5$, filament radius $a=0.1$ \si{\mu m}, and a gap between the cell body and the colloidal sphere of $h=0.01$ \si{\mu m}. The centerline of the left-handed helical flagellum is described by Eq.~\ref{eq:refname13} (Here, the axis of the flagellum is along the $y$-axis direction). In numerical simulation, the spacing between spheres is $s=2.1$. For a stationary colloidal sphere, the corresponding translational and rotational velocities are $U_{s}=0$ and $\Omega_{s}=0$, respectively. The rigid body motion of the cell body is characterized by a translational velocity $\mathbf{U}_{b}$ and a rotational velocity $\boldsymbol{\Omega}_{b}$. Consequently, any arbitrary sphere on the rigid flagellum has the following rotational and translational velocities:
\begin{equation}
\begin{split}
&\Omega_{t}(\mathbf{r})=\Omega_{b}+\Omega_{0}\\
&\mathbf{U}_{t}(\mathbf{r})=\mathbf{U}_{b}+\Omega_{t}(\mathbf{r})\times\left(\mathbf{r}-\mathbf{r}_{b}\right).
\label{eq:refname16}
\end{split}
\end{equation}
where $\Omega_{0}=2\pi f$ is the rotation rate of the motor, and $\mathbf{r}$ and $\mathbf{r}_{b}$ are the position vectors of the centers of the spheres on the flagellum and the cell body, respectively. With the rotation frequency of the motor set at $f=100$ \si{Hz}, we determine the velocities of the cell body, $\mathbf{U}_{b}$ and $\boldsymbol{\Omega}_{b}$, using the linear Eq.~\ref{eq:refname2}. Since a freely moving microswimmer experiences no external forces and torques, we impose the force and torque balance conditions on the microswimmer model system
\begin{equation}
\begin{split}
&\mathbf{F}_{b}+\sum_{i=1}^{N}\mathbf{F}_{t}^{i}=0,\\
&\mathbf{T}_{b}+\sum_{i=1}^{N}\mathbf{T}_{t}^{i}+\sum_{i=1}^{N}\left(\mathbf{r}_{t}^{i}-\mathbf{r}_{b}\right)\times\mathbf{F}_{t}^{i}=0.
\label{eq:refname17}
\end{split}
\end{equation}
where $N$ is the number of spheres on the flagellum, and $\mathbf{F}_{t}$, $\mathbf{F}_{b}$, $\mathbf{T}_{t}$, and $\mathbf{T}_{b}$ are the forces and torques exerted on the fluid by the flagellum and the cell body, respectively.

The resistance matrix of the flagellum can be calculated using both the TMM and RFT. The RFT determines the total force and torque acting on the fluid from the flagellum by integrating the local forces exerted on the fluid by each small segment of the flagellum. The force exerted by a segment of the flagellum on the fluid is expressed as follows:
\begin{equation}
\begin{split}
&d\mathbf{f}=\rm{R}\cdot\mathbf{u}dl.
\label{eq:refname18}
\end{split}
\end{equation}
where $dl$ is the element length and $\mathbf{u}$ is its velocity. The matrix $\rm{R}$ is the local hydrodynamic interaction matrices:
\begin{equation}
\begin{split}
&\rm{R}=k_{\parallel}\hat{\mathbf{t}}\otimes\hat{\mathbf{t}}+k_{\perp}\left(\mathbb{I}-\hat{\mathbf{t}}\otimes\hat{\mathbf{t}}\right).
\label{eq:refname19}
\end{split}
\end{equation}
where $\hat{\mathbf{t}}$ is the local tangential unit vector. The Gray and Hancock's drag coefficients are\cite{Gray1955,Chwang1975}:
\begin{equation}
\begin{split}
&k_{\parallel}=\frac{2\pi\mu}{\ln(2\lambda/a)-1/2},\\
&k_{\perp}=\frac{4\pi\mu}{\ln(2\lambda/a)+1/2}.
\label{eq:refname20}
\end{split}
\end{equation}

\begin{figure}
\centering
   \includegraphics[width=0.48\textwidth]{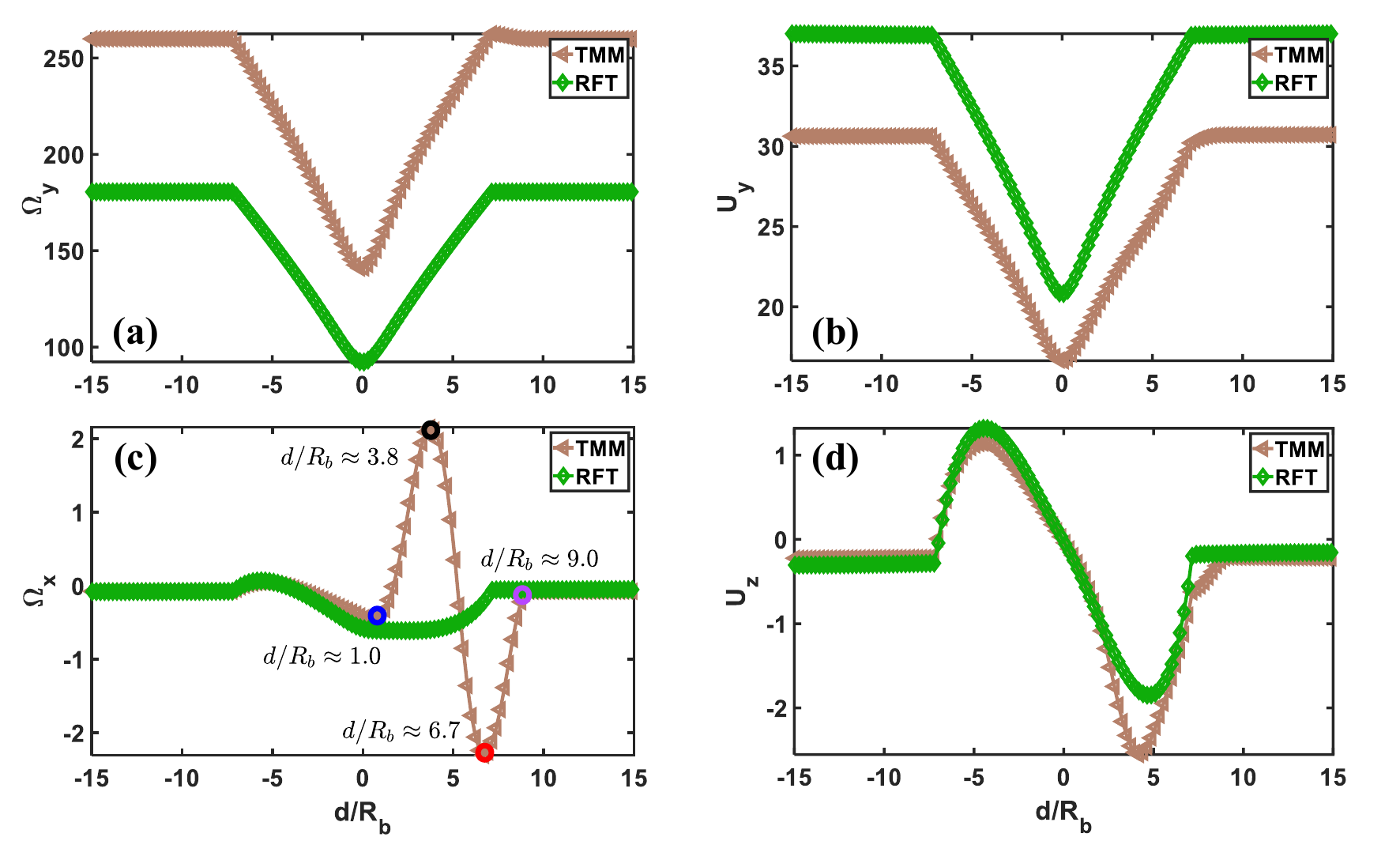}
    \caption{Phase-averaged velocities of the cell body are obtained using the TMM and RFT, respectively. (a)(c) The phase-averaged rotational velocities of the cell body as functions of $d$. (b)(d) The phase-averaged translational velocities of the cell body as functions of $d$.}\label{fig:fig10}
\end{figure}

As illustrated in Fig.~\ref{fig:fig9}, when a microswimmer is far away from the colloidal sphere, we define the rotational resistance matrix of the flagellum along the axial direction as $\zeta^{r}$ and the propulsion matrix of the flagellum as $\zeta^{t}$. By combining our simulation results with the laboratory measurements from Rodenborn \cite{Rodenborn2013}, we observe that when $\lambda/R=2\pi/\tan\theta=8.65$, the matrices of the flagellum obtained using the RFT, LSBT, and TMM satisfy the relationships: $\zeta_{TMM}^{t}\approx\zeta_{LSBT}^{t}\approx\zeta_{RFT}^{t}$ and $\zeta_{TMM}^{r}\approx\zeta_{LSBT}^{r}>\zeta_{RFT}^{r}$. The resistance matrices of the microswimmer model are calculated using both RFT and TMM. Based on the torque balance condition along the $y$-axis for the flagellum and the cell body, we have the condition $\zeta_{TMM}^{r}(\Omega_{0}-\Omega_{y}^{TMM})=\zeta_{RFT}^{r}(\Omega_{0}-\Omega_{y}^{RFT})$, which implies that $\Omega_{y}^{TMM}>\Omega_{y}^{RFT}$. It is evident that the phase-averaged rotational velocity of the cell body along the $y$-axis, obtained using TMM, is greater than that obtained using RFT, as shown in Fig.~\ref{fig:fig10}(a). Furthermore, based on the force balance conditions along the $y$-axis for the flagellum and the cell body, we have: $\zeta_{TMM}^{t}(\Omega_{0}-\Omega_{y}^{TMM})\approx6\pi\mu R_{b}U_{y}^{TMM}$ and $\zeta_{RFT}^{t}(\Omega_{0}-\Omega_{y}^{RFT})\approx6\pi\mu R_{b}U_{y}^{RFT}$. This suggests that $U_{y}^{TMM}<U_{y}^{RFT}$, indicating that the phase-averaged translational velocity along the $y$-axis obtained using TMM is smaller than that obtained using RFT, as depicted in Fig.~\ref{fig:fig10}(b).

When a microswimmer swims near a colloidal sphere, the hydrodynamic interaction between the cell body and the colloidal sphere is calculated using the TMM. Additionally, both TMM and RFT are employed to calculate the resistance matrix of the flagellum. The phase-averaged translational and rotational velocities of the cell body are illustrated in Fig.~\ref{fig:fig10}. The transverse resistance from the colloidal sphere hinders both the translation and rotation of the cell body and flagellum. The relationship between the torque and the rotational velocity of the cell body along the $y$-axis is given by $T_{y}\propto Y_{11}^{C}\Omega_{y}$, where the scalar resistance function $Y_{11}^{C}$ is shown in Fig.~\ref{fig:fig5}. Similarly, the relationship between the force and the translational velocity along the $y$-axis is $F_{y}\propto Y_{11}^{A}U_{y}$, with $Y_{11}^{A}$ displayed in Fig.~\ref{fig:fig2}. As the cell body approaches the colloidal sphere, both $Y_{11}^{C}$ and $Y_{11}^{A}$ increase, leading to reduced phase-averaged velocities, as shown in Fig.~\ref{fig:fig10}(a) and (b). 

Similarly, the transverse resistance from the colloidal sphere causes the cell body to rotate in the negative direction around the $x$-axis, akin to a tire rolling on the ground. When neglecting the hydrodynamic interaction between the flagellum and the colloidal sphere, the relationship between the force along the $y$-axis and the rotational velocity along the $x$-axis of the cell body is given by: $F_{y}\propto Y_{11}^{B}\Omega_{x}$, where the scalar resistance function $Y_{11}^{B}$, shown in Fig.~\ref{fig:fig3}, is a negative value. Therefore, as illustrated in Fig.~\ref{fig:fig10}(c), when a microswimmer swims near the colloidal sphere, the phase-averaged rotational velocity $\Omega_{x}$ is negative, indicating that the microswimmer tends to orient itself towards the direction of the colloidal sphere.

\begin{figure}
\centering
   \includegraphics[width=0.40\textwidth]{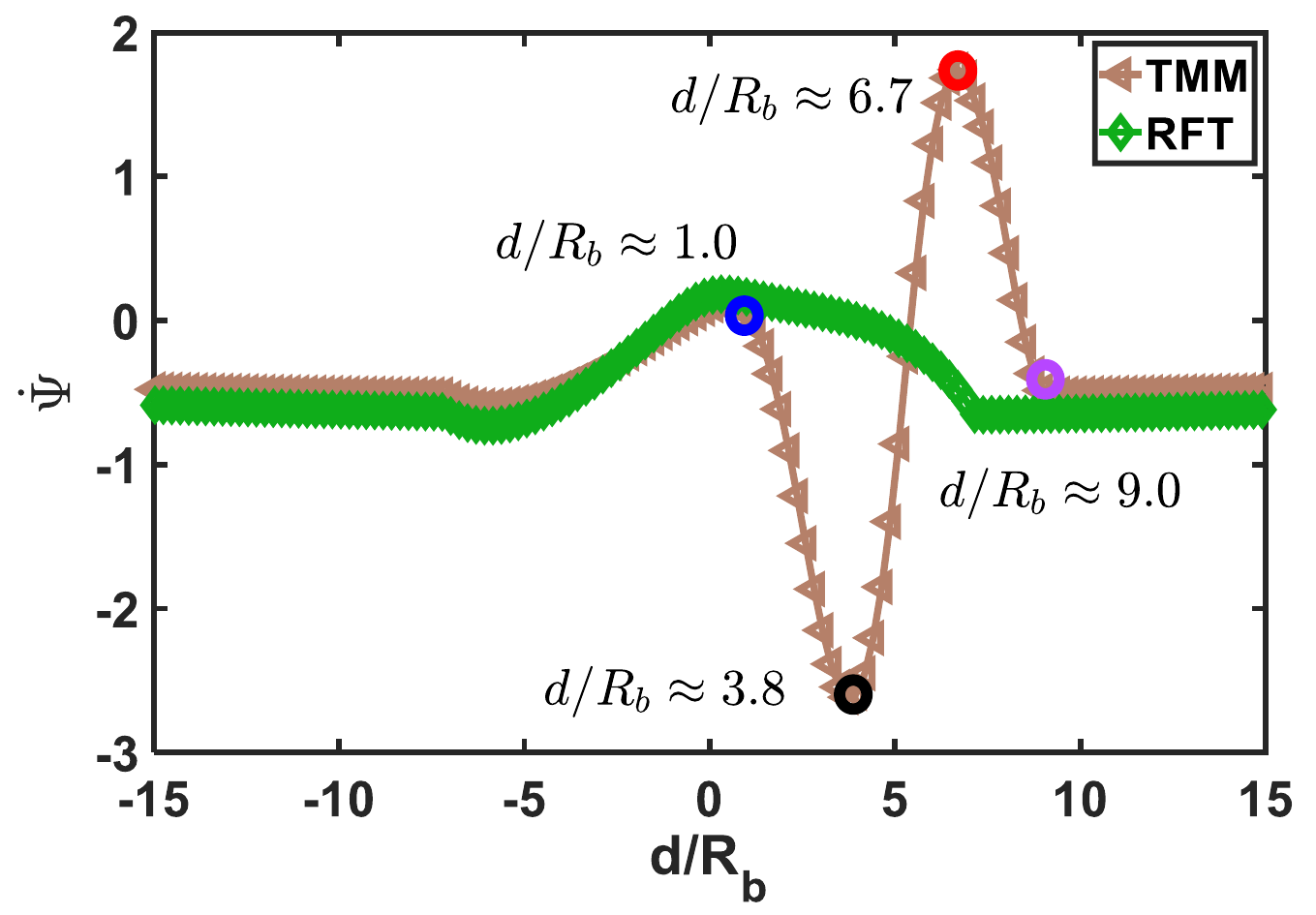}
    \caption{The time derivative of $\Psi$ is obtained using the TMM and RFT, respectively. The circular dots represent the turning points.}\label{fig:fig11}
\end{figure}

The motility patterns of a monotrichous microswimmer in the vicinity of a colloidal sphere is analyzed through phase-plane analysis \cite{Sipos2015,Shum2015,Pimponi2016,Shum2010}:
\begin{equation}
\begin{cases}
\dot{r}(t)=f(r(t),\Psi(t)),\\
\dot{\Psi}(t)=g(r(t),\Psi(t)).
\label{eq:refname21}
\end{cases}
\end{equation}
where $r=\sqrt{(R_{s}+R_{b}+h)^{2}+d^{2}}$ is the distance between the center of the cell body and the colloidal sphere. The quantity $\dot{\Psi}$ characterizes the capture capacity of the colloidal sphere for bacteria. The time derivatives of $r$ and $\Psi$ are expressed as follows:
\begin{equation}
\begin{cases}
\dot{r}(t)=U_{y}\sin\phi+U_{z}\cos\phi,\\
\dot{\Psi}(t)=-\left[\Omega_{x}+\frac{U_{y}\cos\phi-U_{z}\sin\phi}{r}\right].
\label{eq:refname22}
\end{cases}
\end{equation}
where $\sin\phi=d/r$, and $\cos\phi=(R_{s}+R_{b}+h)/r$. 

Given that the quantity $\dot{r}$ primarily depends on the translational velocity $U_{z}$ of the cell body along the $z$-axis, as shown in Fig.~\ref{fig:fig10}(d), the translational velocities $U_{z}$ obtained from the TMM and RFT are remarkably similar. Consequently, we focus exclusively on the variation in $\dot{\Psi}$ in this section. The hydrodynamic interaction between the flagellum and the colloidal sphere significantly influences the velocities of the microswimmer, as illustrated in Fig.~\ref{fig:fig10}(c). In this context, the difference in the time derivatives of $\Psi$ obtained through the TMM and RFT is notable, highlighting the impact of hydrodynamic interaction between the flagellum and the colloidal sphere on the motility patterns of the microswimmer. There are four turning points in Fig.~\ref{fig:fig11}. They correspond to $d\approx R_{b}$, $d\approx R_{b}+L/2$, $d\approx R_{b}+L$, and $d\approx 9R_{b}$. The first turning point indicates that the cell body is moving away from the colloidal sphere, marking the beginning of dominant hydrodynamic interaction between the flagellum and the colloidal sphere. The second point corresponds to the center of the flagellum, $R_{b}+L/2=4$ \si{\mu m}. The third turning point indicates that the microswimmer is moving away from the colloidal sphere. In particular, $d/R_{b}\approx9.0$ indicates that the hydrodynamic interaction between the microswimmer and the colloidal sphere can be considered negligible at this distance. These four turning points are consistent with those shown in Fig.~\ref{fig:fig10}(c). The results shown in Figs.~\ref{fig:fig10}-\ref{fig:fig11} clearly demonstrate that incorporating hydrodynamic interactions among the cell bodies, flagella, and colloidal spheres significantly influences the movement of the microswimmer. Our model effectively incorporates these hydrodynamic interactions, significantly improving the accuracy and efficiency of system simulations.

\section{Summary and Conclusions}
\hangafter=-1\hangindent=19pt\noindent

The TMM proposed by Jeffrey and Onishi does not provide a clear definition of the segmentation point between the near field and the far field when calculating the resistance matrix~\cite{Jeffrey1984}. This ambiguity leads to the creation of independent near- and far-field resistance matrices, which results in a gap in the resistance matrix for the intermediate region. In this work, we define the segmentation point between the near-field and the far-field precisely as $\xi=\min\{\lambda,1/\lambda\}$. We then seamlessly integrate the near- and far-field resistance matrices based on the work of Jeffrey \cite{Jeffrey1984} and Brady \cite{Brady1988} to construct a continuous resistance matrix that covers the entire domain. This approach incorporates both near-field lubrication effects and far-field hydrodynamic interactions among a set of rigid spheres of arbitrary size immersed in a viscous fluid. The far-field mobility matrix, denoted as $\mathcal{M}^{\infty}$, accounts exclusively for long-range hydrodynamic interactions and is mathematically equivalent to the Rotne-Prager Yamakawa (RPY) tensor. Initially, we invert $\mathcal{M}^{\infty}$ to derive a far-field approximation of the grand resistance matrix $\mathcal{R}$. However, this approximation for a multi-sphere system does not include lubrication effects. To address this limitation, we introduce near-field lubrication effects into the grand resistance matrix in a pairwise additive manner, referred to as $\mathcal{R}_{2B,lub}$. It is important to note that lubrication effects are only applicable for $\xi\leqslant\min\{\lambda,1/\lambda\}$. Our comprehensive formulation of the grand resistance matrix $\mathcal{R}$, which incorporates both near-field lubrication effects and far-field hydrodynamic interactions, is expressed as $\mathcal{R}=(\mathcal{M}^{\infty})^{-1}+\mathcal{R}_{2B,lub}$. Once the grand resistance matrix $\mathcal{R}$ is obtained, the grand mobility matrix can be derived by inverting it, yielding $\mathcal{M}=(\mathcal{R})^{-1}$. This technique for constructing the grand resistance matrix offers three main merits: it prevents sphere overlap, approximately satisfies the no-slip boundary condition, and incorporates all long-range interactions.

We consider a rotating, non-translating helical rigid flagellum, modeled as a series of $N$ identical, uniformly spaced spheres. The non-dimensional axial thrust and torque for the flagellum at varying pitches are illustrated in Fig.~\ref{fig:fig8}. The grand resistance matrix constructed in this study captures both near-field lubrication effects and far-field hydrodynamic interactions. To assess the impact of the inter-sphere separation distance, we compute models with varying values of $s$. Our results demonstrate remarkable agreement with Rodenborn's laboratory measurements in the range $2.0001\leqslant s\leqslant2.5$. This suggests that the choice of $s$ for the rigid flagella model within this interval has little effect on the computational outcomes.

Considering and neglecting the hydrodynamic interaction between the flagella and colloidal spheres leads to distinct bacterial velocities, as shown in Figs.~\ref{fig:fig10}-\ref{fig:fig11}. Our model for calculating the grand resistance matrix has proven effective and efficient for investigating bacterial motility near colloidal spheres. The hydrodynamic interactions among the flagella, cell bodies, and colloidal spheres significantly influence bacterial motility. Furthermore, our method can be readily extended to compute the grand resistance or mobility matrix for spheres of arbitrary size. While our approach is accurate and efficient, it remains computationally expensive. Constructing the near-field lubrication resistance matrix requires $O(N^{2})$ operations, while determining the hydrodynamic interactions among multiple spheres involves inverting the mobility matrix within the grand resistance matrix, requiring $O(N^{3})$ operations. This computational demand limits the size of the systems that can be effectively studied. Nevertheless, our technique facilitates the exploration of models that require high precision.

\section*{Acknowledgments} 
\addcontentsline{toc}{section}{Acknowledgments} 
\hangafter=-1\hangindent=19pt\noindent
We acknowledge computational support from Beijing Computational Science Research Center. This work is supported by the National Natural Science Foundation of China(NSFC) umder Grant No. U2230402 and China Postdoctoral Science Foundation under Grant No. 2022M712927.

\section*{Appendix A. The scalar resistance and mobility functions}
\addcontentsline{toc}{section}{Appendix A. The scalar resistance and mobility functions} 
\hangafter=-1\hangindent=19pt\noindent

We follow the notation of Jeffrey and Onishi \cite{Jeffrey1984}, define the variables $s=2r/(a_{\alpha}+a_{\beta})$, $\xi=s-2$ and $\lambda=a_{\beta}/a_{\alpha}$ (the subscripts $\alpha$, and $\beta$ indicate the labels of spheres, which take all values from $1$ to $N$), where $r$ is the center-to-center distance of the inter-sphere and $a_{\alpha}$ is the radius of the sphere $\alpha$. The scalar resistance functions for $\xi\leqslant\min\{\lambda,1/\lambda\}$ are:

1. The scalar resistance functions $X^{a}$
\begin{equation}
\begin{split}
&X_{\alpha\alpha}^{a}=g_{1}(\lambda)\xi^{-1},\\
&X_{\alpha\beta}^{a}=-\frac{2}{1+\lambda}\left[g_{1}(\lambda)\xi^{-1}\right].
\label{eq:refname15}
\end{split}
\end{equation}
where
\begin{equation}
\begin{split}
g_{1}(\lambda)=2\lambda^{2}(1+\lambda)^{-3}.
\label{eq:refname16}
\end{split}
\end{equation}

2. The scalar resistance functions $Y^{a}$
\begin{equation}
\begin{split}
&Y_{\alpha\alpha}^{a}=g_{2}(\lambda)\ln(\xi^{-1}),\\
&Y_{\alpha\beta}^{a}=-\frac{2}{1+\lambda}\left[g_{2}(\lambda)\ln(\xi^{-1})\right].
\label{eq:refname17}
\end{split}
\end{equation}
where
\begin{equation}
\begin{split}
g_{2}(\lambda)=\frac{4}{15}\lambda(2+\lambda+2\lambda^2)(1+\lambda)^{-3}.
\label{eq:refname18}
\end{split}
\end{equation}

3. The scalar resistance functions $Y^{b}$
\begin{equation}
\begin{split}
&Y_{\alpha\alpha}^{b}=g_{2}(\lambda)\ln(\xi^{-1}),\\
&Y_{\alpha\beta}^{b}=-\frac{4}{(1+\lambda)^{2}}\left[g_{2}(\lambda)\ln(\xi^{-1})\right].
\label{eq:refname19}
\end{split}
\end{equation}
where
\begin{equation}
\begin{split}
g_{2}(\lambda)=-\frac{1}{5}\lambda(4+\lambda)(1+\lambda)^{-2}.
\label{eq:refname20}
\end{split}
\end{equation}

4. The scalar resistance functions $X^{c}$
\begin{equation}
\begin{split}
&X_{\alpha\alpha}^{c}=1,\\
&X_{\alpha\beta}^{c}=0.
\label{eq:refname21}
\end{split}
\end{equation}

5. The scalar resistance functions $Y^{c}$
\begin{equation}
\begin{split}
&Y_{\alpha\alpha}^{c}=g_{2}(\lambda)\ln(\xi^{-1}),\\
&Y_{\alpha\beta}^{c}=g_{4}(\lambda)\ln(\xi^{-1}).
\label{eq:refname22}
\end{split}
\end{equation}
where
\begin{equation}
\begin{split}
g_{2}(\lambda)=\frac{2}{5}\lambda(1+\lambda)^{-1},\quad g_{4}(\lambda)=\frac{4}{5}\lambda^{2}(1+\lambda)^{-4}.
\label{eq:refname23}
\end{split}
\end{equation}

The scalar mobility functions are:
\begin{equation}
\begin{split}
&x_{\alpha\alpha}^{a}=1, \qquad x_{\alpha\beta}^{a}=3/2s^{-1}-(2+2\lambda^2)/(1+\lambda)^{2}s^{-3},\\
&y_{\alpha\alpha}^{a}=1, \qquad y_{\alpha\beta}^{a}=3/4s^{-1}+(1+\lambda^{2})/(1+\lambda)^{2}s^{-3},\\
&y_{\alpha\alpha}^{b}=0, \qquad y_{\alpha\beta}^{b}=-1/2s^{-2},\\
&x_{\alpha\alpha}^{c}=1, \qquad x_{\alpha\beta}^{c}=s^{-3},\\
&y_{\alpha\alpha}^{c}=1, \qquad y_{\alpha\beta}^{c}=-1/2s^{-3}.
\label{eq:refname24}
\end{split}
\end{equation}
The sphere labels $\alpha$ and $\beta$ can be interchanged.

\nocite{*}
\bibliography{aipsamp}

\end{document}